\documentclass{article}
\usepackage{fullpage}

\usepackage[utf8]{inputenc}

\usepackage{amsmath,amssymb}

\usepackage{cite}

\usepackage{nameref,hyperref}

\usepackage{authblk}

\usepackage{placeins}

\usepackage{graphicx}
\usepackage{caption}
\usepackage{subcaption}

\usepackage{algorithm,algorithmic}

\usepackage{tikz}
\usetikzlibrary{decorations.pathreplacing}

\usepackage{tabularx}

\begin{document}


\title{\textsl{Gerbil}: A Fast and Memory-Efficient~$k$-mer Counter with GPU-Support\footnote{A short version of this paper will appear in the proceedings of WABI 2016.}}
\author{Marius Erbert}
\author{Steffen Rechner\thanks{Corresponding author.}\ }
\author{Matthias Müller-Hannemann}
\affil{Institute of Computer Science\\ Martin Luther University Halle-Wittenberg, Germany}
\affil{\small{\texttt{\{erbert,rechner,muellerh\}@informatik.uni-halle.de}}}

\date{\today}

\maketitle

\begin{abstract}
\noindent
A basic task in bioinformatics is the counting of~$k$-mers in genome strings. The \emph{$k$-mer counting problem} is to build a histogram of all substrings of length~$k$ in a given genome sequence. We present the open source~$k$-mer counting software \textsl{Gerbil} that has been designed for the efficient counting of~$k$-mers for~$k\geq32$. Given the technology trend towards long reads of next-generation sequencers, support for large~$k$ becomes increasingly important. While existing~$k$-mer counting tools suffer from excessive memory resource consumption or degrading performance for large~$k$, \textsl{Gerbil} is able to efficiently support large~$k$ without much loss of performance. Our software implements a two-disk approach. In the first step, DNA reads are loaded from disk and distributed to temporary files that are stored at a working disk. In a second step, the temporary files are read again, split into~$k$-mers and counted via a hash table approach. In addition, \textsl{Gerbil} can optionally use GPUs to accelerate the counting step. For large~$k$, we outperform state-of-the-art open source~$k$-mer counting tools for large genome data sets.
\end{abstract}

\section{Introduction}

The counting of~$k$-mers in large amounts of reads is a common task in bioinformatics. The problem is to count the occurrences of all~$k$-long substrings in a large amount of sequencing reads. Its most prominent application is de novo assembly of genome sequences. 
Although building a histogram of~$k$-mers seems to be quite a simple task from an algorithmic point of view, it has  attracted a considerably amount of attention in recent years. In fact, the counting of~$k$-mers becomes a challenging problem for large instances, if it is to be both resource- and time-efficient and therefore makes it an interesting object of study for algorithm engineering. Existing tools for~$k$-mer counting are often optimized for~$k<32$ and lack good performance for larger~$k$. Recent advances in technology towards larger read lengths are leading to the quest to cope with values of~$k$ exceeding~32. Studies elaborating on the optimal choice for the value of~$k$ recommend for various applications relatively high values
\cite{Xavier2014, Chikhi01012014}. In particular, working with long sequencing reads helps to improve accuracy and contig assembly (with~$k$ values in the hundreds)~\cite{Sameith-et-al2016}.
In this paper, we develop a tool with a high performance for  such large values of~$k$. 
\subsection{Related Work}

Among the first software tools that succeeded in counting the~$k$-mers of large genome data sets was Jellyfish \cite{marccais2011fast}, which uses a lock-free hash table that allows parallel insertion. In the following years, several tools were published, successively reducing running time and required memory. BFCounter~\cite{Melsted2011} uses Bloom filters for~$k$-mer counting to filter out rarely occurring~$k$-mers stemming from sequencing errors. 
Other tools like DSK \cite{rizk2013dsk} and KMC \cite{Deorowicz2013}  exploit a two-disk architecture and aim at reducing expensive IO operations. Turtle \cite{Roy10032014} replaces a standard Bloom filter by a cache-efficient counterpart. MSPKmerCounter \cite{li2015mspkmercounter} introduces the concept of minimizers to the~$k$-mer counting, thus further optimizing the disk-based approach. The minimizer approach was later on refined to signatures within KMC2 \cite{Deorowicz15052015}. Up to now, the two most efficient open source software tools have been KMC2 and DSK. KMC2 uses a sorting based counting approach that has been optimized for~$k<32$. However, its performance drops when~$k$ grows larger. Instead, DSK uses a single large hash table and is therefore efficient for large~$k$ (but does not support~$k > 127$). However, for small~$k$, it is clearly slower than KMC2.
To the best of our knowledge, the only existing approach that uses GPUs for counting~$k$-mers is the work by Suzuki et al.~\cite{Suzuki2014}.

\subsection{Contribution}
 
In this article we present the open source~$k$-mer counting tool \textsl{Gerbil}. Our software is the result of an extensive process of algorithm engineering that tried to bring together the best ideas from the literature. 
The result is a~$k$-mer counting tool that is both time efficient and memory frugal.\footnote{The tool is named \textsl{Gerbil} because of its modest resource requirements, which it has in common with the name-giving mammal.}
In addition, \textsl{Gerbil} can optionally use GPUs to accelerate the counting step. It outperforms its strongest competitors both in efficiency and resource consumption significantly. For large values of~$k$, it reduces the runtime by up to a factor of four. The software is written in C++ and CUDA and is freely available at \url{https://github.com/uni-halle/gerbil} under MIT license.

In the next section we describe the general algorithmic work flow of \textsl{Gerbil}. Thereafter, in Section~\ref{sec:implementation}, we focus on algorithm engineering aspects that proved essential for high performance and describe details, like the integration of a GPU into the counting process. In Section~\ref{sec:results}, we evaluate \textsl{Gerbil}'s performance in a set of experiments and compare it with those of KMC2 and DSK. We conclude this article by a short summary and a glance on future work.

\section{Methods}

\textsl{Gerbil} is divided into two phases: (1) Distribution and (2) Counting. In this section, we give a high-level description of \textsl{Gerbil}'s work flow.

\subsection{Distribution}
\begin{figure}[t]
	\centering
	\includegraphics[width=0.4\textwidth]{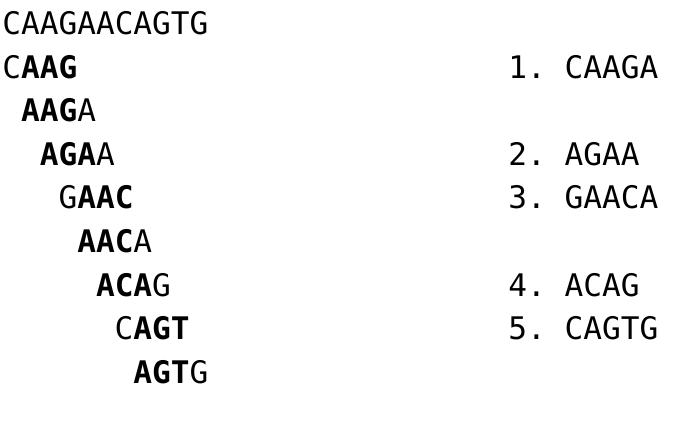}
	\caption{Minimizers and super-mers of the DNA string \texttt{CAAGAACAGTG}. Here,~$k=4$ and~$m=3$. For each~$k$-mer, the bold part is its minimizer. The example uses the lexicographic ordering on 3-mers based on~$A<C<G<T$. The sequence is divided into the five super-mers \texttt{CAAGA}, \texttt{AGAA}, \texttt{GAACA}, \texttt{ACAG}, and \texttt{CAGTG} that would be stored in temporary files.}
	\label{Fig:MiminizerExample}
\end{figure}

Whole genome data sets typically do not fit into the main memory. Hence, it is necessary to split the input data into a couple of smaller temporary files. 
\textsl{Gerbil} uses a two-disk approach that is similar to those of most contemporary~$k$-mer counting tools \cite{Deorowicz15052015, rizk2013dsk, li2015mspkmercounter}. The first disk contains the input read data and is used to store the counted~$k$-mer values. We call this disk input/output-disk. The second disk, which we call working disk, is used to store temporary files. 
The key idea is to assure that the temporary files partition the input reads in such a way, that all occurrences of a certain~$k$-mer are stored in the same temporary file. This way, one can simply count the~$k$-mers of the temporary files independently of each other, with small main memory requirements.
To split the genome data into temporary files, we make use of the \emph{minimizer} approach that has been proposed by~\cite{roberts2004preprocessor} and later on refined by~\cite{Deorowicz15052015}. 
A genome sequence can be decomposed into a number of overlapping \emph{super-mers}. Each super-mer is a substring of maximal length such that all~$k$-mers on that substring share the same minimizer. Hereby, a minimizer of a~$k$-mer is defined as its lexicographically smallest substring of a fixed length~$m < k$ with respect to some total ordering on strings of length~$m$. 
See Fig.~\ref{Fig:MiminizerExample} for an example.
It suffices to partition the set of super-mers into different temporary files to achieve a partitioning of all different~$k$-mers \cite{Deorowicz15052015}.

\subsection{Counting}
The counting of~$k$-mers is typically done by one of two approaches: Sorting and Compressing~\cite{Deorowicz15052015} or using a hash table with~$k$-mers as keys and counters as values~\cite{marccais2011fast, rizk2013dsk}. The efficiency of the sorting approach typically relies on the sorting algorithm Radix Sort, whose running time increases with the length of~$k$-mers. Since we aim at high efficiency for large~$k$, we decided to implement the hash table approach. Therefore, we use a specialized hash table with~$k$-mers as keys and counters as values. We use a hash table that implements open addressing and solves collisions via double hashing. Alg.~\ref{Alg:HashInsert} shows a high level description of the insertion method.

\begin{algorithm}
     \caption{Insert the~$k$-mer~$x$ into the hash table~$t$.} \label{Alg:HashInsert}
     \begin{algorithmic}[1]
       \REQUIRE~Hash Table~$t$,~$k$-mer~$x$, Maximal number of trials~$\theta$. 
       \STATE~$i \gets 0$
       \WHILE[while max number of trials not reached]{$i< \theta$}
       	\STATE~$p \gets hash(x,i)$
       	\IF[matching $k$-mer detected]{$t_p = (x, c)$} 
       		\STATE~$t_p \gets (x, c+1)$
       		\RETURN
       	\ELSIF[empty table entry]{$t_p$ is empty}
       		\STATE~$t_p \gets (x,1)$
       		\RETURN
       	\ELSE[entry locked by another~$k$-mer]
 			\STATE~$i \gets i+1$
       	\ENDIF
       \ENDWHILE
      \STATE Start emergency mechanism.
       \end{algorithmic}
\end{algorithm}

\subsection{Work Flow}
Although the following description of the main process is sequential, all of the steps are interleaved and therefore executed in parallel. This is done by a classical pipeline architecture. Each output of a step makes the input of the next. We use ring buffers to connect the steps of the pipeline. 
Such buffers are specialized for all combinations of single (S)/multiple (M) producers (P) and single (S)/multiple (M) consumers (C). 
The actual number of parallel threads depend on the system and is determined by the software at runtime to achieve optimal memory throughput.

\subsubsection{Phase One: Distribution} 
\begin{figure}[t]
	\centering
	\includegraphics[width=\textwidth]{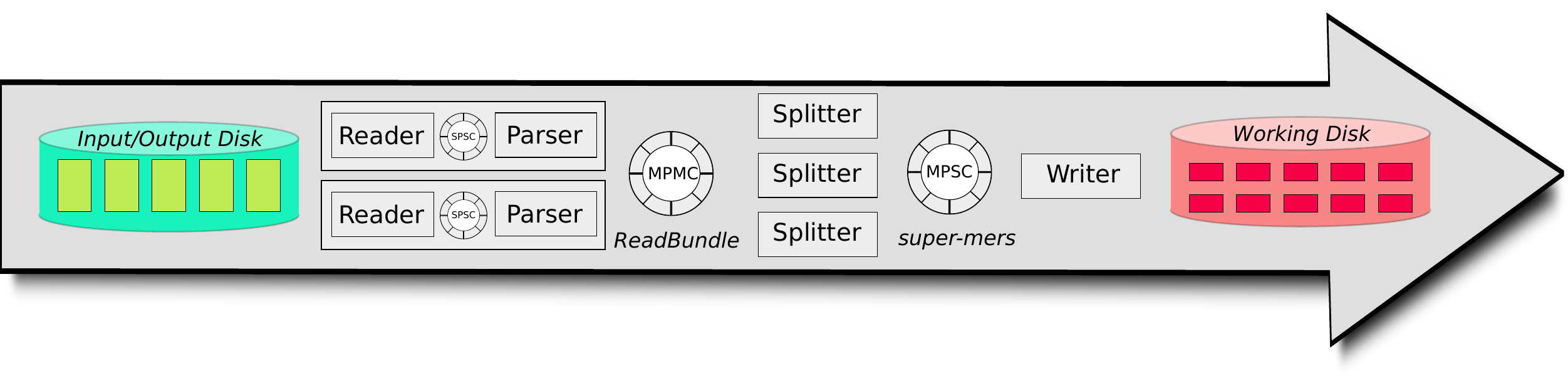}
	\caption{Work flow of Phase One.\label{Fig:Phase1}}
	\includegraphics[width=\textwidth]{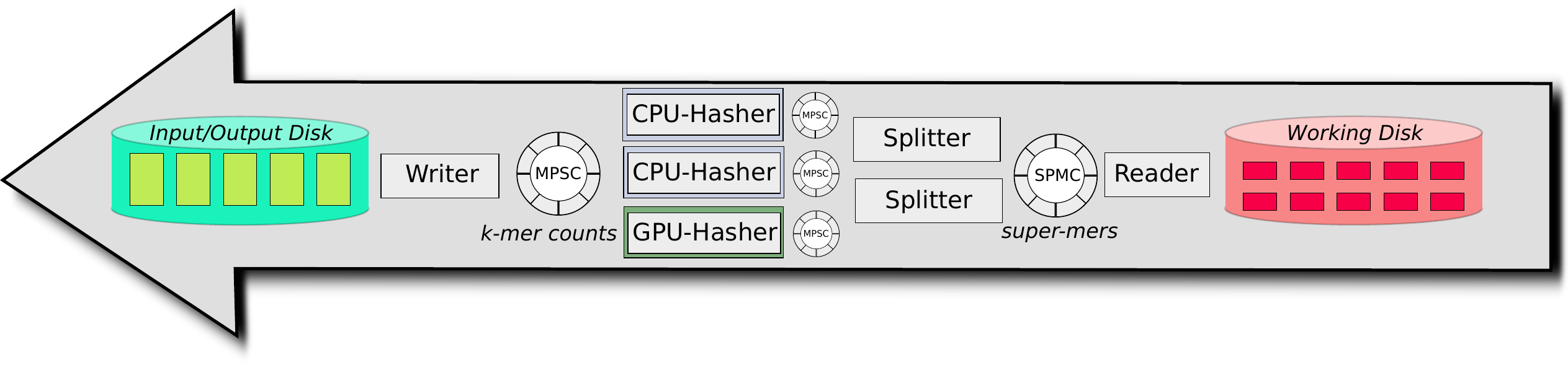}
	\caption{Work flow of Phase Two.\label{Fig:Phase2}}
\end{figure}

The goal of the first phase is to split the input data into a number of temporary files. Fig.~\ref{Fig:Phase1} visualizes the first phase.
\begin{enumerate}
	\item A group of reader threads read the genome reads from the input disk into the main memory. For compressed input, these threads also decompress it.
	\item A second group of parser threads convert the read data from the input format into an internal read bundle format.
	\item A group of splitter threads compute the minimizers of the reads. All subsequent substrings of a read that share the same minimizer are stored as a super-mer into an output buffer.
	\item A single writer thread stores the output buffers to a variable number of temporary files at the working disk.
\end{enumerate}

\subsubsection{Phase Two: Counting}
After the first phase has been completed, the temporary files are sequentially re-read from working disk and processed in the following manner (see Fig.~\ref{Fig:Phase2}).
\begin{enumerate}
	\item A single reader thread reads the super-mers of a temporary file and stores them in main memory.
	\item A group of threads split the super-mers into~$k$-mers. 
	Each~$k$-mer is distributed to one of multiple hasher threads by using a hash function on each~$k$-mer. This ensures that multiple occurrences of the same~$k$-mer are assigned to the same hasher thread and allows the distribution of separated hash tables to different memory spaces.
	\item A group of hasher threads insert the~$k$-mers into their thread-own hash tables. After a temporary file has been completely processed, each hasher thread sends the content of its hash table to an output buffer.
	\item A single writer thread writes from the output buffer to the output disk.
\end{enumerate}

\subsection{DNA Sequence Handling}

\subsubsection{Undetermined bases}
DNA reads typically contain bases that could not been identified correctly during the sequencing process. Usually, such bases are marked $N$ in FASTQ input files. In accordance with established~$k$-mer counting tools, we ignore all~$k$-mers that contain an undetermined base.

\subsubsection{Reverse-Complement}
Since DNA is organized in double helix form, each~$k$-mer~$x \in \{A,C,G,T \}^k$ corresponds to its \emph{reverse-complement} that is defined by reversing~$x$ and replacing~$A \Leftrightarrow T$ and~$C \Leftrightarrow G$. Thus, the~$k$-mer~$ACCG$ corresponds to~$CGGT$. Many applications do not distinguish between a~$k$-mer and its reverse-complement. Thus, each occurrence of~$ACCG$ and~$CGGT$ is counted as occurrences of their unique \emph{canonical} representation. \textsl{Gerbil} uses the lexicographically smaller~$k$-mer as canonical representation. The use of reverse complement normalization can be turned off by command flag.

\section{Implementation Details}\label{sec:implementation}

We now want to point out several details on the algorithm engineering process that were essential to gain high performance.

\subsection{Total ordering on minimizers}

The choice of a total ordering has large effects on the size of temporary files and thus, also on the performance. To find a good total ordering, we have to balance various aspects. On the one hand, the total number of resulting super-mers are to be minimized to reduce the total size of disk memory that is needed by temporary files. On the other hand, the maximal number of distinct~$k$-mers that share the same minimizer should not be too large since we want an approximately uniform distribution of~$k$-mers to the temporary files.
An ``ideal'' total ordering would have both a large total number of super-mers and a small maximal number of distinct~$k$-mers per minimizer. Since these requirements contradict each other, we experimentally evaluated the pros and cons of various ordering strategies.

\begin{description}
\item[CGAT] The lexicographic ordering of minimizers based on~$C < G < A < T$.
\item[Roberts et al. ~\cite{Roberts12122004}] They propose the lexicographic ordering of minimizers with respect to~$C<A<T<G$. Furthermore, within the minimizer computation all bases at even positions are to be replaced by their reverse complement. Thus, rare minimizers like~$CGCGCG$ are preferred.
\item[KMC2] The ordering that is proposed by~\cite{Deorowicz15052015} is a lexicographic ordering with~$A<C<G<T$ and some built-in exceptions to eliminate the large number of minimizers that start with~$AAA$ or~$ACA$.
\item[Random] A random order of all string of fixed length~$m$ is unlikely to have both a small number of super-mers and a highly imbalanced distribution of distinct~$k$-mers. It is simple to establish, since we do not need frequency samples or further assumptions about the distribution of minimizers.
\item[Distance from Pivot (dfp($\mathbf{p}$))] To explain this strategy, consider the following observations: Ascendingly sorting the minimizers by their frequency favors rare minimizers. As a consequence, the maximal number of distinct~$k$-mers per minimizer is small. However, the total number of super-mers can be very large. Similarly, an descendingly sorted ordering results in quite the opposite effect. To find a compromise between both extremes, we initially sort the set of minimizers by their frequency. Since the frequencies depend on the data set, we approximate them by taking samples during runtime. We fix a pivot factor~$0\leq p \leq 1$ and re-sort the minimizers by the absolute difference of their initial position to the pivot position~$4^mp$. The result is an ordering that does neither prefer very rare nor very common minimizers and therefore makes a good compromise.
\end{description}

\begin{figure}[t]
	\centering
	\includegraphics[width=0.5\textwidth]{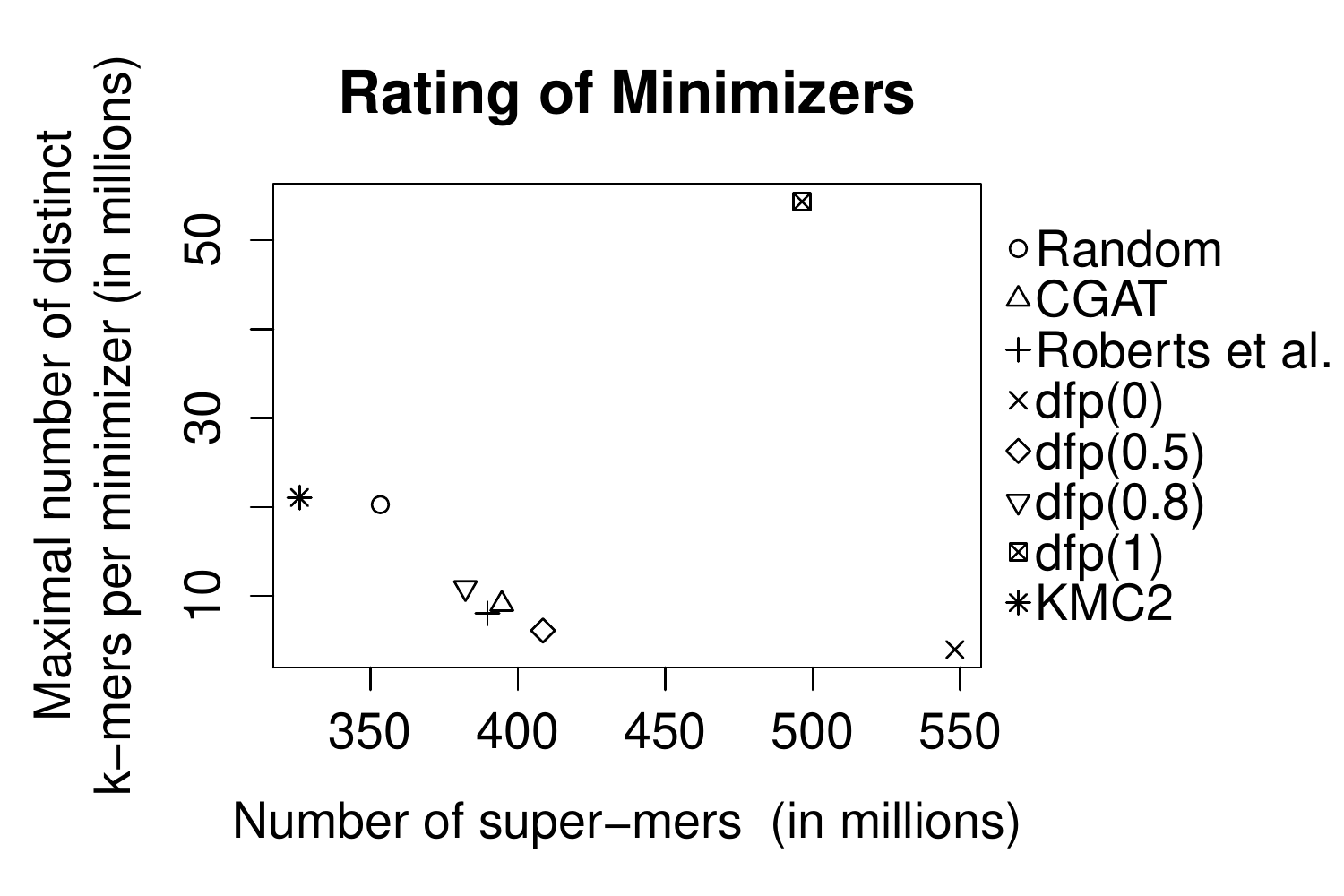}
	\caption{Evaluation of various total ordering strategies for minimizers (\emph{F Vesca},~$m=6$,~$k=28$). Strategy dfp($p$) has been tested with~$p \in \{0, 0.5, 0.8, 1\}$.}
	\label{Fig:Minimizer}
\end{figure}

\subsubsection{Evaluation}
See Fig.~\ref*{Fig:Minimizer} for a rating of each strategy.  The value on the~$x$-axis corresponds to the expected temporary disk memory, whereas the value on the~$y$-axis is correlated with the maximal main memory consumption of our program. A perfect strategy would be located at the bottom left corner. Several strategies seem to be reasonable choices. We evaluated each strategy and found that a small number of super-mers is more important than a small maximal number of~$k$-mer per minimizer for most data sets. As a result, we confirm that the total ordering that is already been used by KMC2 is a good choice for most data sets. Therefore, \textsl{Gerbil} uses the strategy from KMC2 for its ranking of minimizers.

\subsection{Length of miminizers}

The length~$m$ of miminizers is a parameter that has to be chosen with care. However, we can consider a basic rule: The larger $m$ is chosen, the less likely it becomes that consecutive $k$-mers share the same minimizer. Therefore, the number of super-mers decreases with growing $m$. An advantage of a smaller number of super-mers is that the set of super-mers can be distributed to the temporary files more uniformly, which results in temporary files of approximately uniform size. However, a major drawback of a large number of super-mers is the increased total size of temporary files. Thus, a small $m$ results in a better data compression.

In our experiments, we found that choosing minimizer length~$m=7$ is most efficient for the data sets.

\subsection{GPU Integration}

To integrate one or more GPUs into the process of~$k$-mer counting, several problems have to be dealt with. Typically, a GPU performs well only if it deals with data in a parallel manner. In addition, memory bound tasks (i.\,e.\,tasks that do not require a lot of arithmetic operations) like the counting of~$k$-mers require a carefully chosen memory access pattern to minimize the number of the accesses to the GPU's global memory. 
We decided to transfer the hash table based counting approach to the GPU. 

\subsubsection{GPU Hash Tables}
When compiled and executed with GPU support, \textsl{Gerbil} automatically detects CUDA capable GPUs. For each GPU, \textsl{Gerbil} replaces a CPU hasher thread by a GPU hasher thread which maintains its own hash table in GPU memory.
Each GPU hash table is similar in function to a traditional hash table. However, unlike the traditional approach, we add a large number of~$k$-mers in parallel. Therefore, the insertion procedure is slightly changed. 

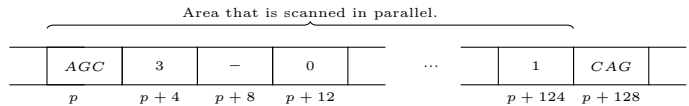
\begin{figure}[t]
	\centering
	\begin{tiny}
		\begin{tikzpicture}
		\draw (0,0.25) -- (1,0.25);
		\draw (0,-0.25) -- (1,-0.25);
		\node [xshift=1cm,draw,rectangle,minimum width=1cm,minimum height=.5cm,label=below:$p\phantom{+0}$] {$AGC$};
		\node [xshift=2cm,draw,rectangle,minimum width=1cm,minimum height=.5cm,label=below:$p+4$] {$3$};
		\node [xshift=3cm,draw,rectangle,minimum width=1cm,minimum height=.5cm,label=below:$p+8$] {$-$};
		\node [xshift=4cm,draw,rectangle,minimum width=1cm,minimum height=.5cm,label=below:$p+12$] {$0$};
		\draw (4.5,0.25) -- (5,0.25);
		\draw (4.5,-0.25) -- (5,-0.25);
		\node [text width=1cm] at (6,0) {...};
		\draw (6,0.25) -- (6.5,0.25);
		\draw (6,-0.25) -- (6.5,-0.25);
		\node [xshift=7cm,draw,rectangle,minimum width=1cm,minimum height=.5cm,label=below:$p+124$] {$1$};
		\node [xshift=8cm,draw,rectangle,minimum width=1cm,minimum height=.5cm,label=below:$p+128$] {$CAG$};
		\draw (8.5,0.25) -- (9,0.25);
		\draw (8.5,-0.25) -- (9,-0.25);
		\draw [decorate, decoration={brace}, yshift=0.5cm] (0.5,0) -- node[above=0.4ex] {Area that is scanned in parallel.}  (7.5,0);
		\end{tikzpicture}
	\end{tiny}
	\caption{GPU memory access pattern. The figure shows the memory area that is being scanned while probing a hash table entry that is stored at memory address~$p$. In this example,~$k=3$ and each table entry needs four bytes for the key and four bytes for the counter. Therefore, 16 entries can be loaded from global memory within one step and are scanned in parallel.}
	\label{Fig:MemAccess}
\end{figure}

First, a bundle of several thousand~$k$-mers is copied to the GPU global memory space. Afterwards, we launch a large number of CUDA blocks, each consisting of 32 threads. Each block sequentially inserts a few~$k$-mers into the GPU hash table.
Since with increasing running time, it becomes more and more probable to find a mismatch when probing a hash table position, we additionally scan adjacent table positions in a range of 128 bytes when probing a hash table entry~(see Fig.~\ref{Fig:MemAccess}). Due to the architecture of a GPU, this can be done within the same global memory access. Thus, we scan up to 16 table entries in parallel, thereby reducing the number of accesses to a GPU's global memory. 
In addition, the total number of probing operations is drastically reduced. In particular, by probing just one entry after the other, 90.37\% of all~$28$-mers of the \emph{F Vesca} data set (Table~\ref{Tab:Data}) can be inserted at first probing and no~$28$-mer needed more than 29 probings. In contrast, through scanning of adjacent table entries, 99.94\% of all~$28$-mers could be inserted at the first trial and no~$28$-mer needed more than seven probing operations.
To eliminate race conditions between CUDA blocks, we synchronize the probing of the hash table by using atomic operations to lock and unlock hash table entries. Since such operations are efficiently implemented in hardware, a large number of CUDA blocks can be executed in parallel.

\subsubsection{Load Balancing} 
We  dynamically balance the amount of~$k$-mers that are assigned to the various CPU and GPU hasher threads. Therefore, we constantly measure the throughput of each hasher thread, i.\,e. the CPU-time needed to insert a certain number of~$k$-mers. Whenever a new temporary file is loaded from disk, we rebalance the number of~$k$-mers that are assigned to each hasher thread, considering the throughput and capacity of each hash table. By that, we automatically determine a good division of labour between CPU and GPU hasher threads without the need of careful hand-tuning.

\subsection{Hash Table Details}

\subsubsection{Estimating table sizes}
\begin{figure}[t]
\centering
\begin{subfigure}[t]{0.45\textwidth}
	\includegraphics[width=\textwidth]{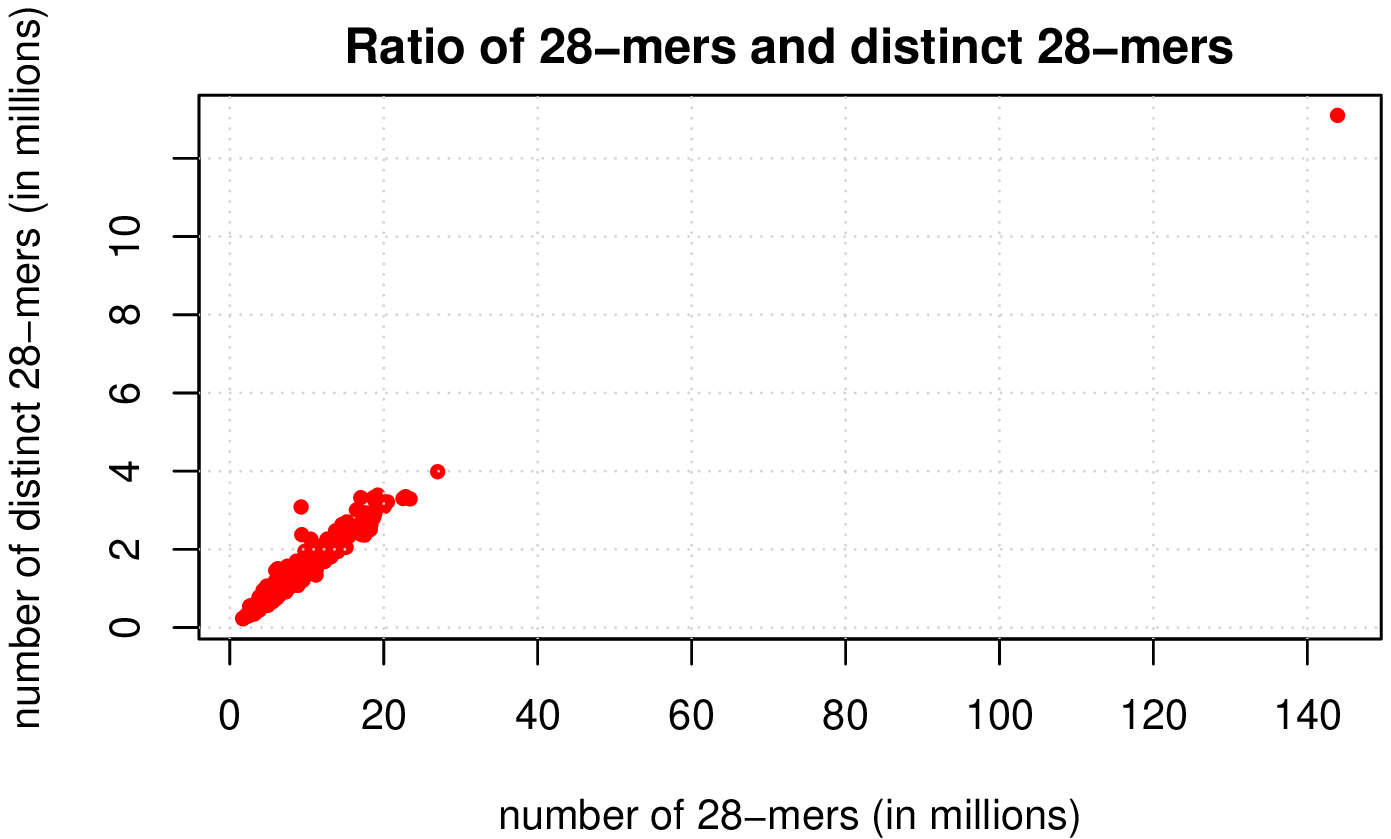}
	\caption{Number of~$28$-mers and number of distinct~$28$-mers in the 512 temporary files that have been created while processing the \emph{F Vesca} data set. Each point corresponds to a temporary file. Here, the KMC2 minimizer ordering did not succeed in creating uniformly sized temporary files since a single file contains far more~$28$-mers than the other 511 files.}
\end{subfigure}
~
\begin{subfigure}[t]{0.45\textwidth}
	\includegraphics[width=\textwidth]{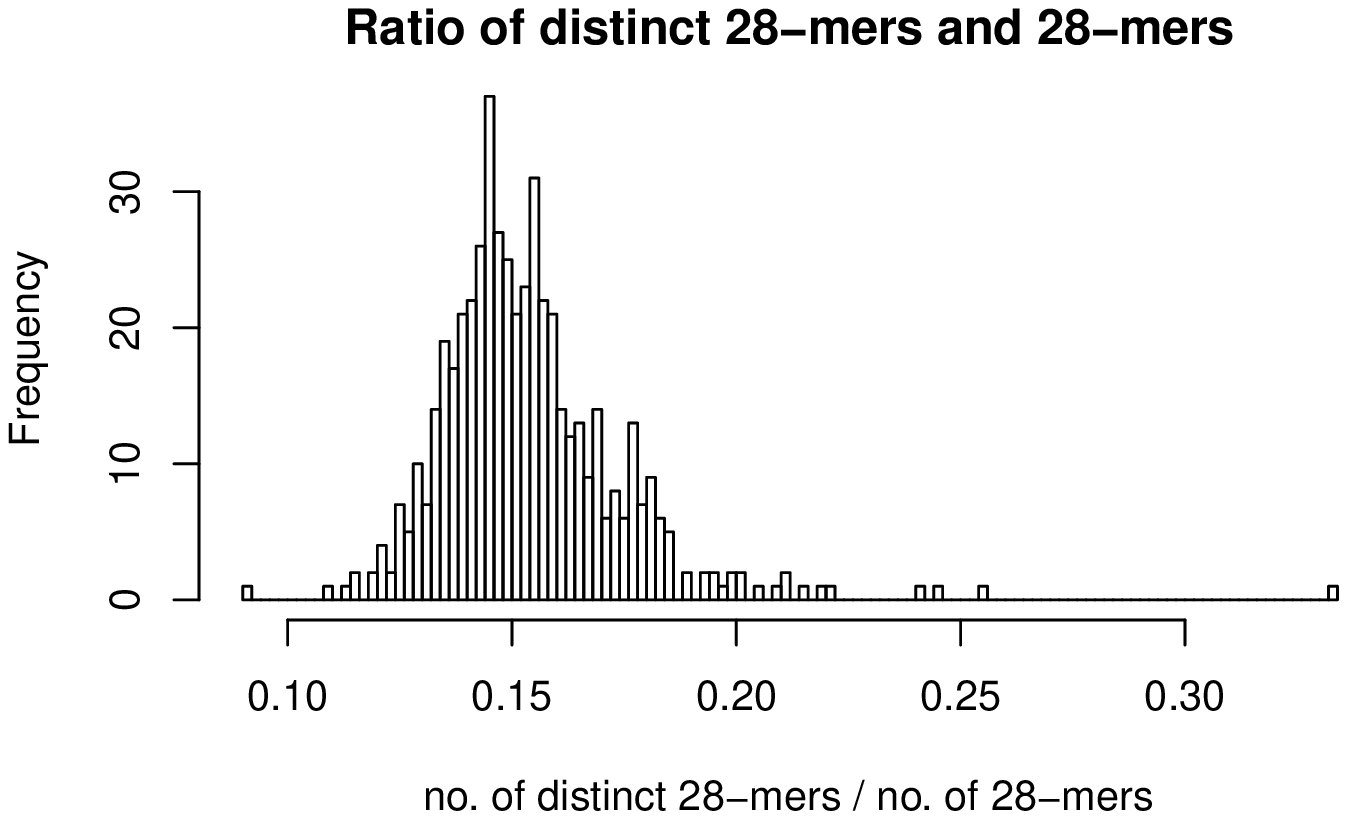}
	\caption{Dividing the number of~$28$-mers in each file by number of its distinct~$28$-mers leads to a ratio that is used to determine the size of the hash tables. A ratio between 0.15 and 0.2 is a proper choice for the \emph{F Vesca} data set.}
\end{subfigure}
	\caption{Estimation of hash table sizes.}
	\label{Fig:KmerUKmers}
\end{figure}

We aim at estimating the expected size of each hash table as closely as possible to save main memory. We do so since reduced memory consumption leaves more memory to the operating system that can be used as cache when writing temporary files. Therefore, we approximate the number of expected distinct~$k$-mers in each temporary file. We use a simple approximation mechanism that predicts the number of distinct~$k$-mers in a file by multiplying the number of~$k$-mers in each file with a constant that has been determined experimentally (see Fig.~\ref{Fig:KmerUKmers}). Since this ratio depends on properties of the data set, we dynamically adjust the ratio during runtime. 

\subsubsection{Hash Function}
To insert a $k$-mer into the hash table, we use a hash function that implements double hashing.

\begin{verbatim}
/**
 * Hash function based on byte-wise interpretation of k-mers.
 * @param kmer: a kmer that shall be inserted
 * @param trial: the number of probings
 */
uint32_t hash(const Kmer<k>& kmer, const uint32_t trial) {
    uint8_t* keyBytes = (uint8_t*) &kmer;   // interpret kmer as byte array
    uint32_t h1 = 0;	// value of first hash function
    uint32_t h2 = 0;	// value of second hash function
    for (uint32_t i = 0; i < sizeof(kmer); i++) {
        h1 = 31 * h1 + keyBytes[i];
        h2 = 37 * h2 + ~keyBytes[i];
    }
    return h1 + trial * h2;
}
\end{verbatim}

\subsubsection{Probing Strategy} As a general strategy we use double hashing. We stop the probing of the hash table after a constant number of trials. Therefore, it is possible that~$k$-mers could not be inserted into a hash table. 
For that reason, \textsl{Gerbil} has a built-in emergency mechanism that handles such~$k$-mers to prevent them from getting lost. Hereby, CPU and GPU hasher threads have different strategies. CPU hasher threads store such~$k$-mers in an additional temporary file, which
is processed after
the work with the current temporary file has been completed. In contrast, GPU hasher threads use part of free GPU memory
    to sequentially store those~$k$-mers that could not be inserted. After all~$k$-mers of a temporary file have been processed, the~$k$-mers in this area are counted via a sorting and compression approach. However, it is still possible to exceed the available GPU memory. In such a case, we copy the whole amount of~$k$-mers in that area back to main memory and store them in a temporary file, similar to the CPU emergency handling. Such an operation is very costly. However, we have never observed a single GPU error handling and only few executions of CPU error handling when processing real world data sets.

\section{Results}\label{sec:results}

\begin{figure}[t]
	\centering
	\begin{subfigure}[b]{0.45\textwidth}
		\includegraphics[width=\textwidth]{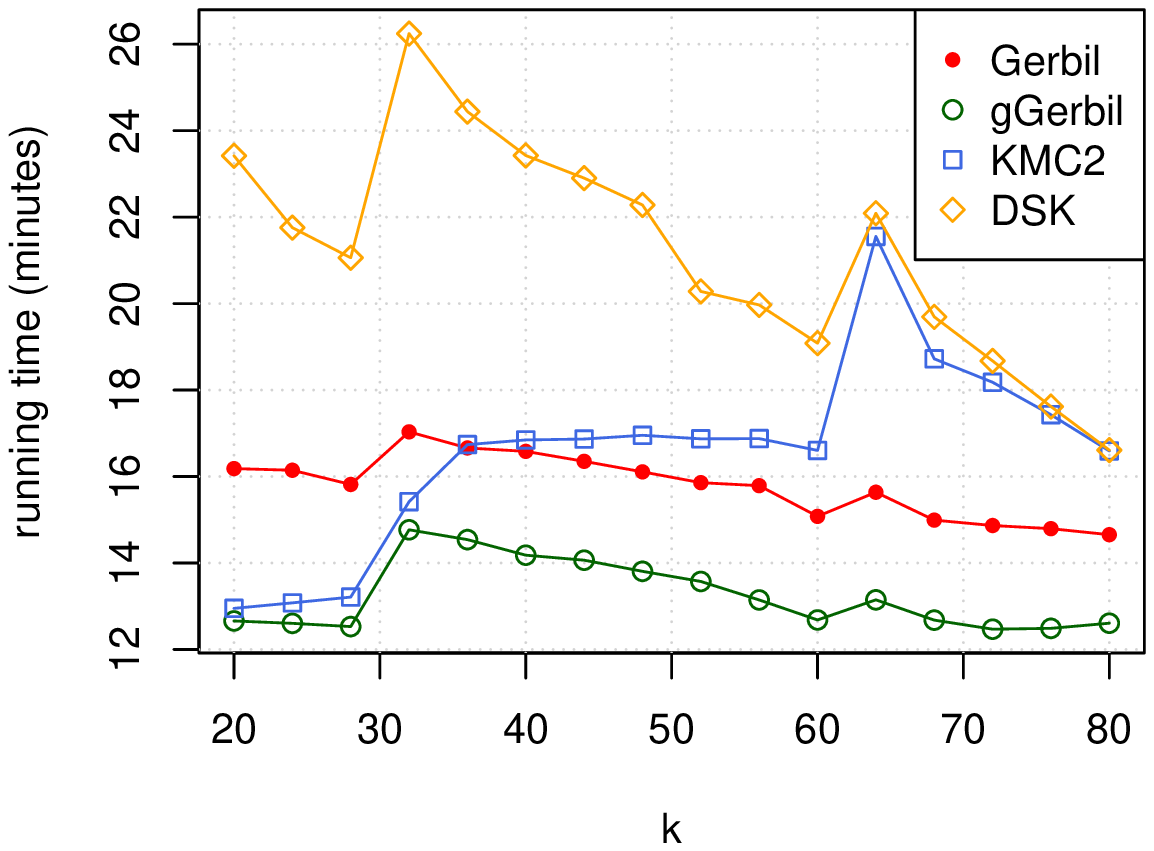}
		\caption{Running times for~$20 \leq k \leq 80$ of \emph{G Gallus} data set.}
		\label{Fig:ScaleK1}
	\end{subfigure}
	~
	\begin{subfigure}[b]{0.45\textwidth}
		\includegraphics[width=\textwidth]{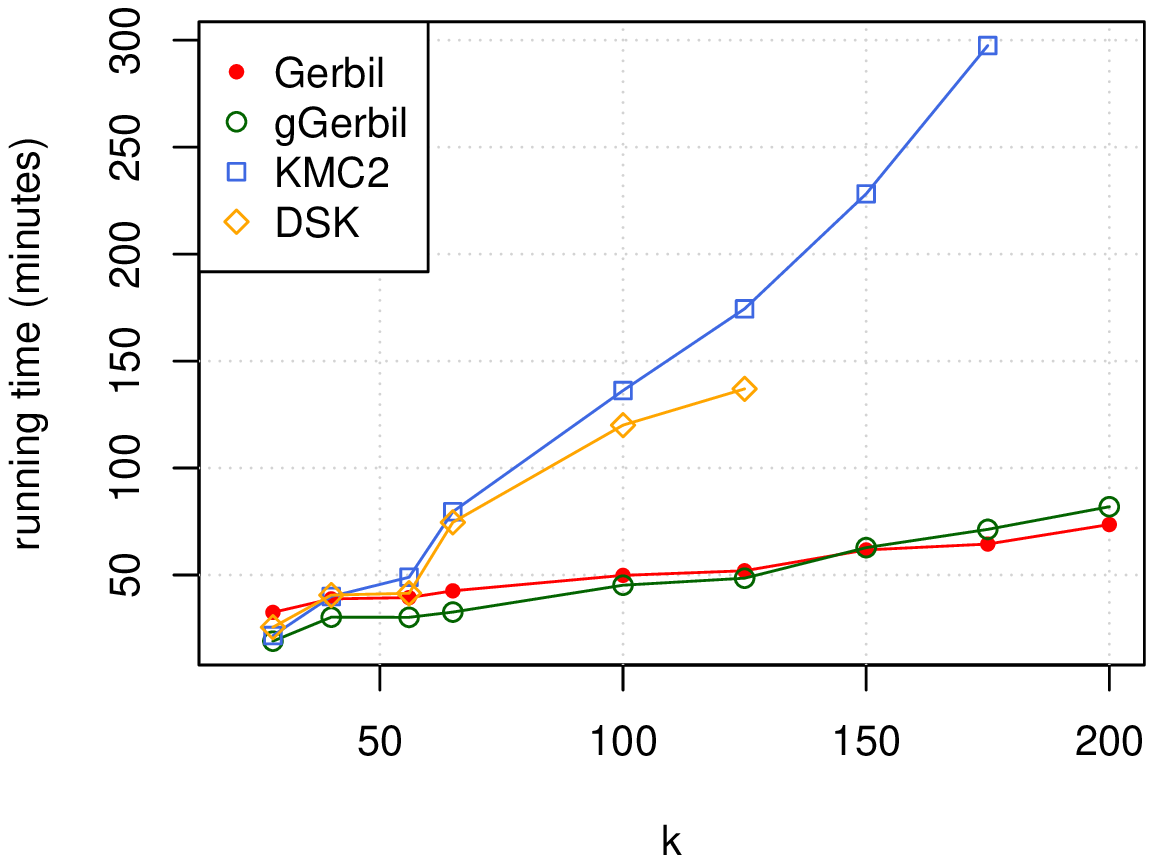}
		\caption{Running times for~$28 \leq k \leq 200$ of \emph{GRCh38} data set.}
		\label{Fig:ScaleK2}
	\end{subfigure}
	\caption{Running time for various $k$. (Test System Two)}
	\label{Fig:ScaleK}	
\end{figure}

We tested our implementation in a set of experiments, using the same instances as Deorowicz et al.~\cite{Deorowicz15052015} 
(see Table~\ref{Tab:Data}). 
For each data set we counted all~$k$-mers for~$k=28,40,56,$ and~$65$ and compared \textsl{Gerbil}'s  running time with those of KMC2 in version 2.3.0 and DSK in version  2.0.7. In addition, we used a synthesized test set \emph{GRCh38}, created from Genome Reference Consortium Human Reference 38 (GCA\_000001405.2), from which we uniformly  
sampled~$k$-mers of size 1000. The purpose of this data set is to 
have longer reads allowing to test the performance for larger values of~$k$.
To judge performance on various types of hardware, we executed the experiments on two different desktop computers. See Table~\ref{Tab:hardware} for details about the hardware configuration of the test systems.
\begin{table}[t]
	\centering
	\caption{Data Sets.}
	\label{Tab:Data}
	\begin{tabular}{lrrrrr}
		\hline\noalign{\smallskip}
		Data Set & Format & Size (GB) & Read Length & $28$-mers & distinct~$28$-mers\\
		\noalign{\smallskip}
		\hline
		\noalign{\smallskip}
		\emph{F Vesca} & FASTQ & 10.2 & 353 & 4 134 078 256 & 632 436 468\\
		\emph{M Balbisiana} & FASTQ & 98.6 & 100 & 20 531 572 597 & 965 691 662 \\
		\emph{G Gallus} & FASTQ & 115.9 & 101 & 25 337 974 831 & 2 727 529 829 \\
		\emph{H Sapiens} & FASTQ & 223.3 & 100 & 62 739 461 708 & 6 336 805 684\\
		\emph{H Sapiens 2} & FASTQ & 339.5 & 101 & 98 892 620 173 & 6 634 382 141\\
		\emph{GRCh38} & FASTA & 100.0 & 1000 & 97 300 000 000 & 1 802 953 276\\
		\hline
	\end{tabular}	
	\centering
	\caption{Test Systems.}
	\label{Tab:hardware}
	\begin{tabular}{lll}
		\hline\noalign{\smallskip}
		& System One & System Two\\
		\noalign{\smallskip}
		\hline
		\noalign{\smallskip}
		CPU & Intel Core-i5 2550k (4 cores) & Intel Xeon(R) E3-1231 v3 (8 cores)\\
		RAM & 16 GB DDR3 & 32 GB DDR3\\
		GPU & GeForce GTX 970 & GeForce GTX TITAN X \\ 
		    & & GeForce GTX 970 \\
		Working-Disk & 256 GB Crucial M550 & 2x Samsung 850 EVO 500 GB (RAID-0)\\
		Free disk space & 128 GB & 1000 GB \\
		OS & \multicolumn{2}{c}{Ubuntu 14.04 LTS \phantom{xxxxxxxxxx}  }\\
		In/Out-Disk & \multicolumn{2}{c}{Transcend StoreJet 35T3 USB 3.0 (External HDD)} \\
		\hline
	\end{tabular}
\end{table}
Table~\ref{Tab:runningTimes} and Fig.~\ref{Fig:ScaleK}  show the results of the performance evaluation.
We want to point out several interesting observations.
\begin{itemize}
	\item \textsl{Gerbil} with GPU support (\textsl{gGerbil}) is the most efficient tool in almost all cases. Exceptions occur for small~$k=28$, where the sorting based approach KMC2 is sometimes slightly more efficient.
	\item Fig.~\ref{Fig:ScaleK1} shows that Gerbil starts outperforming KMC2 at $k\approx36$. Interestingly, the running time of each tool decreases with growing $k$. This can be explained by the small read length of the \emph{G Gallus} data set. In addition, one can observe the erratic increase of running time near~$k=32$ and~$k=64$ for all tools, due to a change of the internal~$k$-mer representation. 
	\item When~$k$ grows, KMC2 becomes more and more inefficient, while \textsl{Gerbil} stays efficient. 
	When counting the~$200$-mers in the \emph{GRCh38} data set, KMC2 did not finish within 20 hours, whereas \textsl{Gerbil} required only 98 minutes (Fig.~\ref{Fig:ScaleK2}). The running time of DSK grows similarly fast as that of KMC2. Recall that DSK does not support values of~$k > 127$. 
	\item For small~$k$, the use of a GPU decreases the running time by a significant amount of time. However, with growing~$k$, the data structure that stores~$k$-mers grows larger. Therefore, the number of table entries that can be scanned in parallel decreases. Experimentally, we found that the GPU induced speedup vanishes when~$k$ exceeds $150$.
\end{itemize}
\begin{table}[t]
	\centering
	\caption{Running times in the format mm:ss (the best performing in bold). Each entry is the average over three runs. Missing running times for DSK are due to insufficient disk space. The label `gGerbil' stands for \textsl{Gerbil} with activated GPU mode. Instead, standard `Gerbil' does not use any GPU.}
	\label{Tab:runningTimes}
	\begin{footnotesize}
	\begin{tabular}{ll|rrrc|rrrc}
		\hline
		\noalign{\smallskip}
		\multicolumn{2}{l}{} & \multicolumn{4}{c}{\textbf{System One}} & \multicolumn{4}{c}{\textbf{System Two}}\\
		Data Set &~$k$ & Gerbil & {gGerbil} & {KMC2} & {DSK} & Gerbil & {gGerbil} & {KMC2} & {DSK}\\
		\noalign{\smallskip}
		\hline
		\noalign{\smallskip}
		\emph{F Vesca} & 28 & 02:08 & {\bf 01:40} & 02:01 & 03:00 & 01:36 & {\bf01:18} & 01:32 & 02:05 \\
		& 40 & 02:34 & {\bf 01:53} & 03:03 & 04:14 & 02:01 & {\bf01:38} & 02:12 & 02:52 \\
		& 56 & 02:58 & {\bf 01:53} & 03:19 & 03:55 & 02:25 & {\bf01:39} & 02:30 & 02:50 \\
		& 65 & 03:05 & {\bf 01:59} & 04:34 & 05:23 & 02:16 & {\bf01:42} & 03:35 & 03:37 \\
		\noalign{\smallskip}
		\hline
		\noalign{\smallskip}
		\emph{M Balbisiana} & 28 & 13:37 & {\bf 11:42} & 12:54 & 14:49 & 11:17 & {\bf10:07} & 10:50 & 11:06 \\
		& 40 & 13:48 & {\bf 12:24} & 16:15 & 16:12 & 11:46 & {\bf10:59} & 13:46 & 12:26 \\
		& 56 & 12:46 & {\bf 11:36} & 16:06 & 14:56 & 10:50 & {\bf 10:18} & 13:36 & 11:44 \\
		& 65 & 12:32 & {\bf 11:28} & 18:33 & 15:52 & 10:46 & {\bf10:16} & 15:47 & 12:34 \\
		\noalign{\smallskip}
		\hline
		\noalign{\smallskip}
		\emph{G Gallus} & 28 & 18:41 & {\bf 14:25} & 15:39 & 26:54 & 15:47 & {\bf12:31} & 13:10 & 21:00 \\
		& 40 & 19:55 & {\bf 16:00} & 19:44 & 29:42 & 16:29 & {\bf14:10} & 16:49 & 23:48\\
		& 56 & 18:12 & {\bf 14:48} & 19:48 & 24:11 & 15:38 & {\bf13:12} & 16:48 & 19:59 \\
		& 65 & 18:27 & {\bf 15:22} & 22:49 & 26:50 & 15:41 & {\bf13:08} & 19:25 & 21:33 \\		
		\noalign{\smallskip}
		\hline
		\noalign{\smallskip}
		\emph{H Sapiens} & 28 & 41:10 & {\bf 30:04} & 32:18 & - & 33:26 & {\bf25:16} & 26:44 & 50:15 \\
		& 40 & 45:02 & {\bf 35:52} & 43:19 & - & 35:20 & {\bf29:00} & 35:59 & 54:21 \\
		& 56 & 39:47 & {\bf 33:21} & 42:53 & - & 32:21 & {\bf26:46} & 35:25 & 45:32 \\
		& 65 & 38:09 & {\bf 35:32} & 51:23 & - & 32:09 & {\bf26:27} & 42:19 & 47:50 \\
		\noalign{\smallskip}
		\hline
		\noalign{\smallskip}
		\emph{H Sapiens 2} & 28  & 65:33 & 49:41 & {\bf 49:17} & - & 53:40 & {\bf39:24} & 41:47 & 76:50  \\
		& 40 & 72:06 & {\bf 66:04} & 70:33 & - & 57:03 & {\bf46:00} & 57:02 & 83:59 \\
		& 56 & 64:00 & {\bf 60:27} & 69:58  & - & 51:34 & {\bf42:15} & 56:28 & 72:35 \\
		& 65 & {\bf 61:05} & 64:44 & 87:24 & - & 51:16 & {\bf41:30} & 68:10 & 78:13 \\
		\hline
	\end{tabular}
	\end{footnotesize}
\end{table}

\noindent
We gain some additional interesting insights when we take a closer look into Table~\ref{Tab:runningTimes2} that shows detailed information on running time and memory usage.
\begin{itemize}
	\item The use of a GPU accelerates \textsl{Gerbil}'s second phase by up to a factor of two, whereas the additional speedup given by a second GPU is only moderate. 
	\item All tools were called with an option that sets the maximal memory size to 14~GB on Test System One and 30 GB on Test System Two. However, \textsl{Gerbil} typically uses much less memory due to its dynamic prediction of the hash table size. In contrast, both KMC2 and DSK use more main memory. 
	\item \textsl{Gerbil}'s disk usage is comparable to KMC2's disk usage, whereas the disk usage of DSK is much larger. 
	\item \textsl{Gerbil}'s frugal use of disk- \emph{and} main memory is a main reason for its high performance. The use of little main memory gives the operating system opportunity to use the remaining main memory for buffering disk operations. A small disk space consumption is essential since disk operations are far more expensive than the actual counting.
	\item We can compare the actual running time with the theoretically minimal running time that is given by hardware constraints (hereby done with Test System Two). We find that the running time of phase one is bounded by the read rate of the input disk. Thus, a further speedup of the first phase can only be achieved by reading compressed input files. In contrast, there is potential speedup in phase two: In GPU mode, Gerbil's throughput in phase two is just about 61\% of the possible throughput given by write rate of the output disk. In non-GPU mode it is about 35\%. Interestingly, the throughput of the working disk is not a critical component at our test system.
\end{itemize}

\begin{table}[t]
	\centering
	\caption{Detailed running times (in format mm:ss) and maximal main memory and disk space consumption (in GB) for the \emph{G Gallus} instance. Each entry is the average of three runs.}
	\label{Tab:runningTimes2}
	\begin{footnotesize}
	\begin{tabular}{ll|rrrr|rrrr}
		\hline
		\noalign{\smallskip}
		\multicolumn{2}{l}{} & \multicolumn{4}{c}{\textbf{System One}} & \multicolumn{4}{c}{\textbf{System Two}}\\
		& $k$& Gerbil & {gGerbil} & {KMC} & {DSK} & Gerbil & {gGerbil} &  {KMC} & {DSK}\\
		 \noalign{\smallskip}
		\hline
		\noalign{\smallskip}
		Phase 1 & 28 & 10:08 & 10:06 & 10:51 & 10:22 & 09:46 & 09:43 & 09:52 & 09:30 \\
		Phase 2 & 28 & 08:32 & 04:19 & 04:46 & 16:00 & 06:01 & 02:47 & 03:16 & 11:01 \\
		Main Memory & 28 & 2.36 & 1.77 & 14.28 & 15.28 & 2.20 & 2.01 & 26.99 & 16.69 \\
		Disk Space & 28 & 23.66 & 23.66 & 24.86 & 37.30 & 23.66 & 23.66 & 24.86 & 37.30 \\
		\noalign{\smallskip}
		\hline
		\noalign{\smallskip}
		 Phase 1 & 56 & 10:07 & 10:06 & 10:40 & 10:26 & 09:47 & 09:43 & 09:47 & 09:30 \\
		 Phase 2 & 56 & 08:05 & 04:42 & 09:08 & 13:13 & 05:50 & 03:28 & 06:59 & 10:00 \\
		 Main Memory & 56 & 4.24 & 3.20 & 14.29 & 15.00 & 4.00 & 3.40 & 26.98 & 14.78 \\
		 Disk Space & 56 & 16.25 & 16.25 & 17.02 & 57.20 & 16.25 & 16.25 & 17.02 & 57.20 \\
		\hline 
	\end{tabular}
	\end{footnotesize}
\end{table}

\section{Conclusion}

We introduced the~$k$-mer counting software \textsl{Gerbil} that uses a hash table based approach for the counting of~$k$-mers. For large~$k$, a use case that becomes important for long reads, we are able to clearly outperform the state-of-the-art open source~$k$-mer counting tools, while using significantly less resources.
We showed that \textsl{Gerbil}'s running time can be accelerated by the use of GPUs. However, since this only affects the second phase, the overall additional speedup is only very moderate.
As future work, we plan to evaluate strategies to use GPUs to accelerate also the first phase. Another option for further speed-up would be to give up exactness by using Bloom filters.

\FloatBarrier

\clearpage
\begin{appendix}
\section*{Appendix}

\section{Command Line Parameters}
\textsl{Gerbil} can be controlled by several command line arguments and flags.

\begin{flushleft}
\begin{tabularx}{\textwidth}{lXr}
\hline
\noalign{\smallskip}
Command & Description & Default \\
\noalign{\smallskip}
\hline
\noalign{\smallskip}
\verb|-k length| & Set the value of~$k$, i.\,e. the length of~$k$-mers to be counted. Supported~$k$ range from~$8$ to~$479$. & 28\\
\noalign{\smallskip}
\hline
\noalign{\smallskip}
\verb|-m length| & Set the length~$m$ of minimizers. & auto\\
\noalign{\smallskip}
\hline
\noalign{\smallskip}
\verb|-e size| & Restrict the maximal size of main memory in \verb|MB| or \verb|GB| that \textsl{Gerbil} is allowed to use.  & auto \\
\noalign{\smallskip}
\hline
\noalign{\smallskip}
\verb|-f number| & Set the number of temporary files. & 512 \\
\noalign{\smallskip}
\hline
\noalign{\smallskip}
\verb|-t number| & Set the maximal number of parallel threads to use. & auto \\
\noalign{\smallskip}
\hline
\noalign{\smallskip}
\verb|-l count| & Set the minimal occurrence of a~$k$-mer to be outputted.  & 3 \\
\noalign{\smallskip}
\hline
\noalign{\smallskip}
\verb|-i| & Enable additional output. & \\
\noalign{\smallskip}
\hline
\noalign{\smallskip}
\verb|-g| & Enable GPU mode. \textsl{Gerbil} will automatically detect CUDA-capable devices and will use them for counting in the second phase. & \\
\noalign{\smallskip}
\hline
\noalign{\smallskip}
\verb|-v| & Show version number. & \\
\noalign{\smallskip}
\hline
\noalign{\smallskip}
\verb|-d| & Disable normalization of~$k$-mers. If normalization is disabled, a~$k$-mer and its reverse complement are considered as different~$k$-mers. If normalization is enabled, we map both~$k$-mer and its reverse complement to the same~$k$-mer. & \\
\noalign{\smallskip}
\hline
\noalign{\smallskip}
\verb|-s| & Perform a system check and display information about your system. & \\
\noalign{\smallskip}
\hline
\noalign{\smallskip}
\verb|-x 1| &  Stop execution after Phase One. Do not remove temporary files and a \texttt{binStatFile} (with statistical information). Watch out: no \texttt{output} allowed. & \\
\noalign{\smallskip}
\hline
\noalign{\smallskip}
\verb|-x 2| & Only execute Phase Two. Requires temporary files and the \texttt{binStatFile}. & \\
\noalign{\smallskip}
\hline
\noalign{\smallskip}
\verb|-x b| & Do not remove the \texttt{binStatFile}. & \\
\noalign{\smallskip}
\hline
\noalign{\smallskip}
\verb|-x h| & Create a histogram of $k$-mers in a human readable format in output directory. & \\
\hline
\end{tabularx}
\end{flushleft}

\section{Input Formats}

\textsl{Gerbil} supports the following input formats of genome read data: FASTQ, FASTA, staden, as well as compressed files of these formats. To process multiple files, it also supports textfiles with paths to one or more input files.

\section{Output Format}

\textsl{Gerbil} uses an output format that is easy to parse and requires little space. The counter of each occuring~$k$-mer is stored in binary form, followed by the corresponding byte-encoded~$k$-mer. Each four bases of a~$k$-mer are encoded in one single byte. We encode A with 00, C with 01, G with 10 and T with 11. Most counters of~$k$-meres are slightly smaller than the coverage of the genome data. We exploit this property by using only one byte for counters less than 255. A counter greater than or equal to 255 is encoded in five bytes. In the latter case, all bits of the first byte are set to 1. The remaining four bytes contain the counter in a conventional 32-bit unsigned integer.
Examples (X is undefined): 
\begin{itemize}
\item \verb|67 AACGTG|~$\Rightarrow$ \verb|01000011 00000110 1110XXXX|
\item \verb|345 TGGATC|~$\Rightarrow$ \verb|11111111 00000000 00000000 00000001 01011001 11101000 1101XXXX|
\end{itemize}

When called with command line argument \verb|-x h|, \textsl{Gerbil} additionally creates a human readable csv file including the same, but uncompressed information. This option should only be used for very small data sets.

\section{Datasets}

The data sets were downloaded from the following URL's.

\paragraph{F Vesca} ~\\
\begin{footnotesize}
\url{ftp://ftp.ddbj.nig.ac.jp/ddbj_database/dra/fastq/SRA020/SRA020125/SRX030575/SRR072005.fastq.bz2}\\
\url{ftp://ftp.ddbj.nig.ac.jp/ddbj_database/dra/fastq/SRA020/SRA020125/SRX030576/SRR072006.fastq.bz2}\\
\url{ftp://ftp.ddbj.nig.ac.jp/ddbj_database/dra/fastq/SRA020/SRA020125/SRX030576/SRR072007.fastq.bz2}\\
\url{ftp://ftp.ddbj.nig.ac.jp/ddbj_database/dra/fastq/SRA020/SRA020125/SRX030577/SRR072008.fastq.bz2}\\
\url{ftp://ftp.ddbj.nig.ac.jp/ddbj_database/dra/fastq/SRA020/SRA020125/SRX030577/SRR072009.fastq.bz2}\\
\url{ftp://ftp.ddbj.nig.ac.jp/ddbj_database/dra/fastq/SRA020/SRA020125/SRX030575/SRR072010.fastq.bz2}\\
\url{ftp://ftp.ddbj.nig.ac.jp/ddbj_database/dra/fastq/SRA020/SRA020125/SRX030575/SRR072011.fastq.bz2}\\
\url{ftp://ftp.ddbj.nig.ac.jp/ddbj_database/dra/fastq/SRA020/SRA020125/SRX030575/SRR072012.fastq.bz2}\\
\url{ftp://ftp.ddbj.nig.ac.jp/ddbj_database/dra/fastq/SRA020/SRA020125/SRX030578/SRR072013.fastq.bz2}\\
\url{ftp://ftp.ddbj.nig.ac.jp/ddbj_database/dra/fastq/SRA020/SRA020125/SRX030578/SRR072014.fastq.bz2}\\
\url{ftp://ftp.ddbj.nig.ac.jp/ddbj_database/dra/fastq/SRA020/SRA020125/SRX030578/SRR072029.fastq.bz2}\\
\end{footnotesize}

\paragraph{G Gallus} ~\\
\begin{footnotesize}
\url{ftp://ftp.ddbj.nig.ac.jp/ddbj_database/dra/fastq/SRA030/SRA030308/SRX043656/SRR105788_1.fastq.bz2}\\
\url{ftp://ftp.ddbj.nig.ac.jp/ddbj_database/dra/fastq/SRA030/SRA030308/SRX043656/SRR105788_2.fastq.bz2}\\
\url{ftp://ftp.ddbj.nig.ac.jp/ddbj_database/dra/fastq/SRA030/SRA030309/SRX043656/SRR105789_1.fastq.bz2}\\
\url{ftp://ftp.ddbj.nig.ac.jp/ddbj_database/dra/fastq/SRA030/SRA030309/SRX043656/SRR105789_2.fastq.bz2}\\
\url{ftp://ftp.ddbj.nig.ac.jp/ddbj_database/dra/fastq/SRA030/SRA030312/SRX043656/SRR105792_1.fastq.bz2}\\
\url{ftp://ftp.ddbj.nig.ac.jp/ddbj_database/dra/fastq/SRA030/SRA030312/SRX043656/SRR105792_2.fastq.bz2}\\
\url{ftp://ftp.ddbj.nig.ac.jp/ddbj_database/dra/fastq/SRA030/SRA030314/SRX043656/SRR105794.fastq.bz2}\\
\url{ftp://ftp.ddbj.nig.ac.jp/ddbj_database/dra/fastq/SRA030/SRA030314/SRX043656/SRR105794_1.fastq.bz2}\\
\url{ftp://ftp.ddbj.nig.ac.jp/ddbj_database/dra/fastq/SRA030/SRA030314/SRX043656/SRR105794_2.fastq.bz2}\\
\url{ftp://ftp.ddbj.nig.ac.jp/ddbj_database/dra/fastq/SRA036/SRA036382/SRX043656/SRR197985.fastq.bz2}\\
\url{ftp://ftp.ddbj.nig.ac.jp/ddbj_database/dra/fastq/SRA036/SRA036382/SRX043656/SRR197985_1.fastq.bz2}\\
\url{ftp://ftp.ddbj.nig.ac.jp/ddbj_database/dra/fastq/SRA036/SRA036382/SRX043656/SRR197985_2.fastq.bz2}\\
\url{ftp://ftp.ddbj.nig.ac.jp/ddbj_database/dra/fastq/SRA036/SRA036383/SRX043656/SRR197986.fastq.bz2}\\
\url{ftp://ftp.ddbj.nig.ac.jp/ddbj_database/dra/fastq/SRA036/SRA036383/SRX043656/SRR197986_1.fastq.bz2}\\
\url{ftp://ftp.ddbj.nig.ac.jp/ddbj_database/dra/fastq/SRA036/SRA036383/SRX043656/SRR197986_2.fastq.bz2}\\
\end{footnotesize}

\paragraph{M Balbisiana} ~\\
\begin{footnotesize}
\url{ftp://ftp.ddbj.nig.ac.jp/ddbj_database/dra/fastq/SRA098/SRA098922/SRX339427/SRR956987.fastq.bz2}\\
\end{footnotesize}

\paragraph{H Sapiens} ~\\
\begin{footnotesize}
\url{ftp://ftp.1000genomes.ebi.ac.uk/vol1/ftp/phase3/data/HG02057/sequence_read/SRR359301.filt.fastq.gz}\\
\url{ftp://ftp.1000genomes.ebi.ac.uk/vol1/ftp/phase3/data/HG02057/sequence_read/SRR359301_1.filt.fastq.gz}\\
\url{ftp://ftp.1000genomes.ebi.ac.uk/vol1/ftp/phase3/data/HG02057/sequence_read/SRR359301_2.filt.fastq.gz}\\
\url{ftp://ftp.1000genomes.ebi.ac.uk/vol1/ftp/phase3/data/HG02057/sequence_read/SRR360755.filt.fastq.gz}\\
\url{ftp://ftp.1000genomes.ebi.ac.uk/vol1/ftp/phase3/data/HG02057/sequence_read/SRR360755_1.filt.fastq.gz}\\
\url{ftp://ftp.1000genomes.ebi.ac.uk/vol1/ftp/phase3/data/HG02057/sequence_read/SRR360755_2.filt.fastq.gz}\\
\end{footnotesize}

\paragraph{H Sapiens 2} ~\\
\begin{footnotesize}
\url{http://www.ebi.ac.uk/ena/data/view/ERA015743/}\\
\end{footnotesize}

\paragraph{GRCh38} ~\\
\begin{footnotesize}
\url{http://hgdownload.cse.ucsc.edu/goldenPath/hg38/bigZips/hg38.fa.gz}
\end{footnotesize}

\end{appendix}

\end{document}